\documentclass{INTERSPEECH2023}
\usepackage{multirow}

\interspeechcameraready

\title{North S\'{a}mi Dialect Identification with Self-supervised Speech Models}
\name{Sofoklis Kakouros$^1$ and Katri Hiovain-Asikainen$^2$}
\address{
  $^1$University of Helsinki, Finland\\
  $^2$UiT The Arctic University of Norway}
\email{sofoklis.kakouros@helsinki.fi, katri.hiovain-asikainen@uit.no}

\begin{document}

\maketitle
 
\begin{abstract}
The North S\'{a}mi (NS) language encapsulates four primary dialectal variants that are related but that also have differences in their phonology, morphology, and vocabulary. The unique geopolitical location of NS speakers means that in many cases they are bilingual in S\'{a}mi as well as in the dominant state language: Norwegian, Swedish, or Finnish. This enables us to study the NS variants both with respect to the spoken state language and their acoustic characteristics. In this paper, we investigate an extensive set of acoustic features, including MFCCs and prosodic features, as well as state-of-the-art self-supervised representations, namely, XLS-R, WavLM, and HuBERT, for the automatic detection of the four NS variants. In addition, we examine how the majority state language is reflected in the dialects. Our results show that NS dialects are influenced by the state language and that the four dialects are separable, reaching high classification accuracy, especially with the XLS-R model.
%The North S\'{a}mi (NS) language encapsulates four primary dialectal variants that are related but that also have differences in their phonology, morphology, and vocabulary. The unique geopolitical location of NS speakers means that in many cases the speakers are bilingual in S\'{a}mi as well as in the dominant state language: Norwegian, Swedish, and Finnish. This enables us to study the NS variants both with respect to the spoken state language and their acoustic characteristics. In this paper, we investigate an extensive set of standard acoustic features, including prosodic features and MFCCs, as well as state-of-the-art self-supervised representations, namely, XLS-R and HuBERT, for the automatic detection of the four NS variants. In addition, we examine how the majority state language is reflected in the dialects. Looking at the effect of majority language, we find that NS dialects are influenced by the state language. When classifying the dialects, we obtain good performance, reaching high accuracy, especially with the XLS-R model.
\end{abstract}

\section{Introduction}

Dialects are language variants that enfold the language characteristics of a specific region or regional community. Dialects may vary significantly in terms of vocabulary, pronunciation, and grammar, which can pose barriers in the communication between speakers from different regions. Dialect identification (DID) is an important task that has received increasing attention in many areas including automatic speech recognition (ASR), machine translation, and text-to-speech synthesis (TTS). As different dialects may have different patterns of text and speech this in turn affects the performance in these tasks. For example, a machine translation system trained on one dialect may not perform as well on another dialect, leading to errors and misunderstandings. DID is a challenging task as dialectal differences are typically more subtle than differences between languages. This challenge is exacerbated in under-resourced languages by the accessibility to only limited amount of data for the language and more so, for the dialects of interest. Thus, with the increasing access to knowledge, tools, and highly resourced data to build artificial intelligence (AI) systems, it is conversely important to evaluate the portability of these AI systems to under-resourced languages, but also to develop them further.

In this work, we conduct a thorough examination of different acoustic features (prosodic features, MFCCs, filter banks) in NS DID and also perform one of the first studies examining the applicability of self-supervised speech models (XLS-R, WavLM, and HuBERT) on DID, in general, and NS DID, in particular. In addition we examine the impact of the state language on North S\'{a}mi by running a language identification (LID) experiment on the data. We evaluate our method on a collection of five different North S\'{a}mi datasets and show impact of the state language in the speakers’ productions as well as good classification results using XLS-R. The code for our experiments is publicly available at \url{https://github.com/skakouros/sami_dialects}.

\vspace{-2mm}
\subsection{Dialect identification}

State-of-the-art DID solutions have been built using various machine learning approaches in different domains of application (see, e.g., \cite{chiang2017cross} for TTS, \cite{weninger2019deep, liu2011systematic} for ASR) as DID systems can be used to improve their performance. DID can be generally approached based on the text of the dialects \cite{jauhiainen2022italian}, based on the speech \cite{kakouros2020dialect}, or with a combination of both \cite{imaizumi2022end}. For textual resources, the goal of DID is to determine if a sentence contains dialectal elements (e.g., differences in vocabulary and grammar that are specific to a particular dialect) or if the text belongs to a particular dialect \cite{jauhiainen2022italian, zaidan2014arabic}. For DID with speech, the corresponding objective is to determine whether an utterance is spoken in a manner belonging to a specific dialect \cite{kakouros2020dialect, wang2021end}. In the case of fusion approaches, text and speech are combined to determine the dialect \cite{imaizumi2022end}. In this work, we focus solely on detecting the dialect using representations extracted from speech.

DID from speech can be performed using various machine learning methods as well as using different combinations of acoustic features (see, e.g., \cite{ali2016automatic,wang2021end,kakouros2020dialect,Shon2018}). In \cite{simko2018mumbling} Slavic dialects were compared using Continuous Wavelet Transform (CWT) and mutual perplexity measures based on their $f_0$ and energy envelope signals. In \cite{suni2019comparative}, a deep neural network approach commonly used in speech synthesis based on WaveNet modeling \cite{oord2016wavenet} was applied to identify relationships between tonal varieties of Swedish and evaluate differences in prosodic features across regions. The results showed that the method can produce a meaningful clustering of the varieties based solely on prosodic information, agreeing with previous analyses of Swedish tonal accents and their regional variation (see, e. g. \cite{gaarding1977scandinavian}).

%Using prosodic features to cluster dialects and languages has been shown to be an effective way of measuring prosodic distance and similarity between spoken language varieties. \cite{cummins1999automatic} was one of the first attempts to use machine learning for complex statistical modeling of speech. LSTM (Long short-term memory) networks on $f_0$ and energy contours were trained to measure inter-language distance across five prosodically different languages (English, Japanese, Spanish, Mandarin Chinese and German). The discrimination task between different languages based on local changes in $f_0$ and amplitude envelope was typologically meaningful and significantly better than chance. 

%Following a similar approach, \cite{simko2017comparing} used a data-driven approach to assess the differences in prosodic characteristics across seven European languages. Using Continuous Wavelet Transform and unigram language models, the languages were compared using mutual perplexity measures. Groupings of the languages based on their $f_0$ and energy envelope signals demonstrated a possible ``prosodic distance'' between the languages while also suggesting patterns of language contacts and influence between neighboring countries. Similar methodology was successful also in clustering a number of different Slavic language varieties, as shown in \cite{simko2018mumbling}. The results of the clustering indicated that the prosodic characteristics of minority languages could be influenced by the majority language.

\vspace{-2mm}
\subsection{The North S\'{a}mi language and its dialects}

\begin{figure}[tb]
  \centering
  \includegraphics[width=\linewidth]{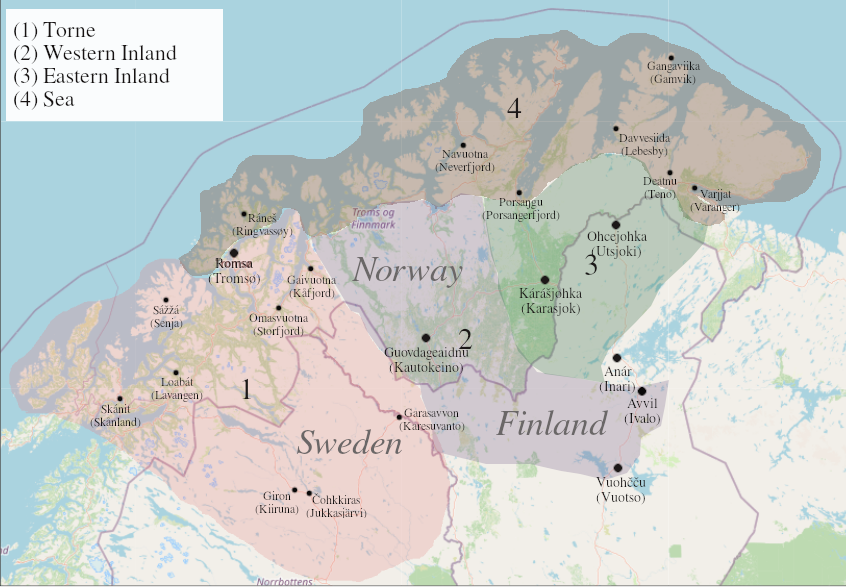}
  \caption{Map of traditional speaking areas and four main dialect groups of North S\'{a}mi. Marked villages indicate the recording locations. Map base (c) OpenStreetMap, data is licensed under the Open Data Commons Open Database License (ODbL). Dialect areas adapted from \cite{aikio2022north}.}
  \label{fig:dialects}
  \vspace{-4mm}
\end{figure}

The North S\'{a}mi language is traditionally spoken in the northernmost parts of Norway, Sweden and Finland (see Figure~\ref{fig:dialects}) \cite{aikio2022north}. With estimated 20,000 -- 30,000 speakers, it is the most widely spoken S\'{a}mi language. Today, due to modern mobility and urbanization, a considerable number of its speakers live outside the traditional speaking area, especially in urban centres such as the capital areas of the three countries.

According to the traditional dialectological analysis of North S\'{a}mi (see \cite{aikio2022north, palismaa2001gielas, sammallahti1998saami}), the language can be divided to four main dialect groups: the Western Inland, the Eastern Inland\footnote{\label{footnote:inland}Often, the `Inland' dialects are called the `Finnmark' dialects, according to the Troms og Finnmark county, which is the northernmost county of Norway. This is somewhat misleading since the same dialects of North S\'{a}mi are also spoken in the northernmost parts of Finland. Here, however, we use the name ``Finnmark''.}, the Torne, and the Sea, see Figure~\ref{fig:dialects}. The differences between the traditional regional dialects are most prominent in phonology, and to a lesser degree, morphology and vocabulary. 

%\begin{figure}[t]
%  \centering
%  \includegraphics[width=\linewidth]{Sami_languages_with_box.png}
%  \caption{A map of northern Scandinavia and Russian far-northeast, where the S\'{a}mi languages are traditionally spoken. The North S\'{a}mi traditional speaking area is marked with the label `Nord'. Map (c) Wikimedia Commons. The blue rectangle indicates roughly the area shown in Figure~\ref{fig:dialects}.}
%  \label{fig:sami_lgs}
%\end{figure}

The unique geopolitical position of North S\'{a}mi is described in \cite{aikio2022north, aikio2015variation}. Currently, most speakers are bilingual in S\'{a}mi and in the dominant state language, and the simultaneous influence of Norwegian, Swedish, and Finnish on different parts of the speaking area has led to the emergence of ``new'' dialect boundaries that coincide with state borders. This has resulted in dialectal variation of more recent origin which can be explained with the influence of the different majority and state languages in the region. This influence has presumably affected many features in the language, including phonetics, syntax, vocabulary and prosody \cite{kakouros2020dialect, kakouros2018comparison, kakouros2017evaluation}.

\vspace{-2mm}

%TODO: add also other works outside S\'{a}mi that have used similar methods to cluster dialects?

% THIS LIST IS PROVISIONAL - pending final version from Kate.
% \setlist{noitemsep,topsep=0pt,parsep=2pt,partopsep=0pt,leftmargin=1em}

\vspace{-2mm}
\section{Related work}

The areal variation of spoken North S\'{a}mi has been explored in a number of papers in the recent years. While the methodologies of these papers vary, all of them have used data-driven approaches. Focusing on the spoken varieties of North S\'{a}mi, in \cite{hiovain2018mapping} and \cite{hiovain2020comparative} it was found that based on prosodic features ($f_0$ and energy envelope) alone, utterances from five areal varieties were clustered following the majority language influence (Norwegian and Finnish) instead of the traditional dialectal groups. When looking at the intonational characteristics of the spoken areal varieties of North S\'{a}mi from the two countries, similarities with the respective majority languages were detected from the North S\'{a}mi intonational contours extracted from the speech materials.

Using somewhat different methodology to analyze partly the same spoken materials, \cite{jokinen2016variation} used i-vector techniques to classify North S\'{a}mi varieties based on acoustic, segmental, and phonotactic features, also focusing on the influence of the predominant majority language. The results supported their hypothesis, as well as the results of \cite{hiovain2018mapping} and \cite{hiovain2020comparative}, suggesting that the variation in spoken North S\'{a}mi varieties is due to the majority language spoken in the given context by bilingual speakers, rather than individual variation.

In addition, \cite{kakouros2020dialect} investigated the influence of purely prosodic features, such as energy, $f_0$, spectral tilt and duration, on the North S\'{a}mi varieties. Using the same dataset as \cite{jokinen2016variation}, it was found that prosodic information played a crucial role in distinguishing between five areal varieties of North S\'{a}mi. Three varieties from Finland and two from Norway were shown to cluster well together, supporting the results of \cite{jokinen2016variation} and \cite{hiovain2018mapping}. The recordings of the data were done at Guovdageaidnu and K\'{a}r\'{a}\v{s}johka in Norway and Avvil, An\'{a}r and Ohcejohka situated in the Western and Eastern Inland dialect areas, marked in Figure~\ref{fig:dialects}. In summary, these methodologies provide valuable insights into the influence of majority languages on North S\'{a}mi areal varieties, particularly in terms of their prosodic features.

However, in the papers mentioned above, only two dialectal groups were included in the analyses: the Western and Eastern Finnmark groups (WF and EF). In the current paper, we include all four dialect groups (WF, EF, TS and SS) to our analysis. With a speech corpus from altogether 147 speakers our aim is to see how the speakers cluster based on an extensive set of acoustic features and self-supervised representations and 1) the dialect group 2) the majority language.

\vspace{-2mm}
\section{Method}

\begin{figure}[tb]
  \centering
  \includegraphics[width=\linewidth]{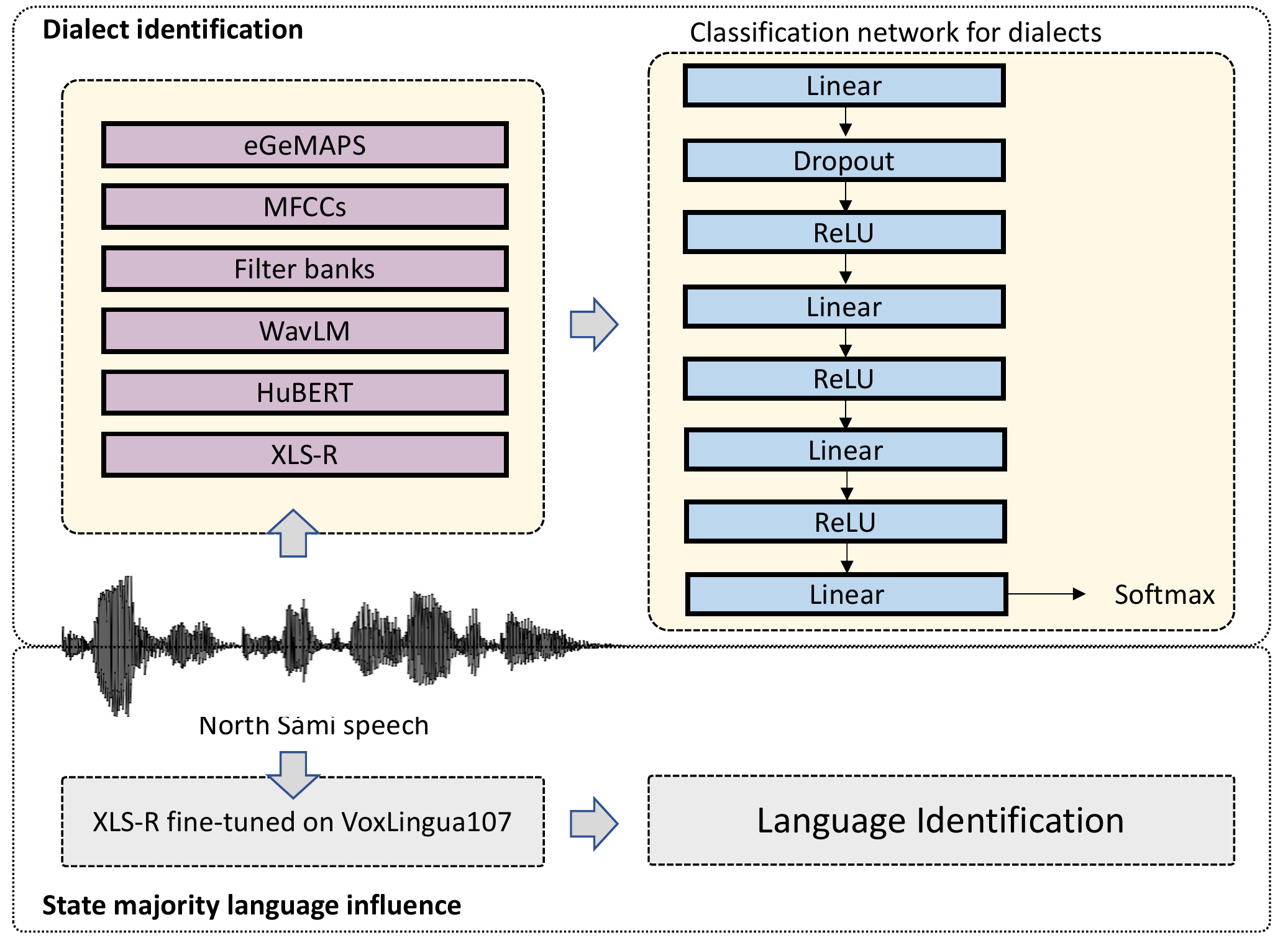}
  \caption{Overview of the experimental setup.}
  \label{fig:overview_exp_setup}
  \vspace{-4mm}
\end{figure}

We use several acoustic features and self-supervised representations in a classification task that includes a light-weight classification network. In addition, we examine the majority language influence. In the next we present details of our methodology.

\vspace{-2mm}
\subsection{Feature extraction}
We extract the following acoustic features: eGeMAPS (88 features including their functionals) \cite{eyben2015geneva}, kaldi Mel-frequency cepstral coefficients (MFCCs; 39 coefficients including deltas and deltadeltas) \cite{povey2011kaldi}, and kaldi filter banks (FBs; 40 bins). In addition, we extract self-supervised representations from three different models: HuBERT \cite{hsu2021hubert}, WavLM \cite{chen2022wavlm}, and XLS-R \cite{babu2021xls} -- features from each of the models is of dimensionality $1024$. For all self-supervised models, we freeze the pre-trained model and take the representation from the last transformer layer as our feature per frame. Dialect identification is an utterance-level task, therefore, we need to pool our frame-level representations to obtain a single vector per utterance/recording. For all frame-level features, we take the mean and standard deviation of all frames across an utterance. This results in three utterance-level descriptors: (i) mean, (ii) standard deviation (std), and (iii) mean + standard deviation (meanstd; mean and std concatenated to a single vector).

\vspace{-2mm}
\subsection{Classification}
For the classification we use a light-weight feed-forward deep neural network (DNN). An overview of the classification setup can be seen in Figure~ \ref{fig:overview_exp_setup}. The network consists of a projection layer to 256, and an additional three hidden layers; the overall network layout is the following $L = [256, 128, 64, 32]$, where $L$ is the network layer. We train the network for 50 epochs, with learning rate $lr=1e-4$ using the adam optimizer. We randomly shuffle the training data at each epoch and feed it into the network in batches of 100. We use a cross-entropy loss function to optimize the network and apply dropout regularization with $p=0.1$ only after the first hidden layer. The best performing model based on the validation accuracy is saved and used to evaluate the test set.

\vspace{-2mm}
\subsection{Majority language influence}
To investigate how the majority language of the region where the North S\'{a}mi dialects are spoken may affect speech, we conduct an experiment on language identification (LID). We take a spoken LID model that is fine-tuned using weights of the pre-trained XLS-R (\url{facebook/wav2vec2-xls-r-300m}) on the VoxLingua107 corpus \cite{valk2021voxlingua107} and use the interface developed by SpeechBrain \cite{speechbrain} for the task. The LID model has been trained on 107 languages including Finnish, Norwegian, and Swedish. Our aim is to examine whether S\'{a}mi dialects are influenced by the majority language of the region, and if that is the case, our hypothesis is that, this should be reflected on the performance of the LID classifier.

\vspace{-2mm}
\section{Experiments}

In the next subsections we describe our data, pre-processing procedure, splits generation, and experimental setup.

\vspace{-2mm}
\begin{figure}[b]
  \centering
  \includegraphics[width=\linewidth]{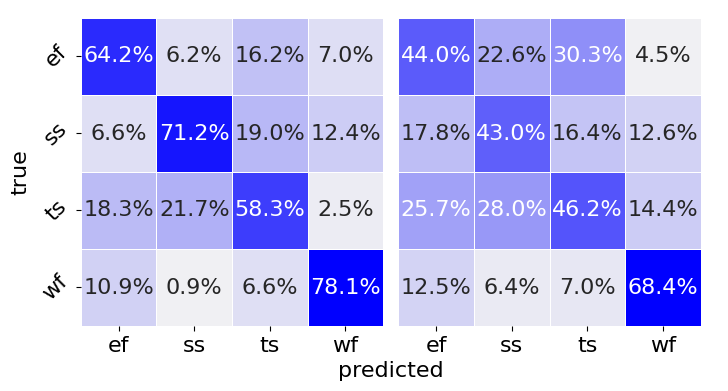}
  \caption{Confusion matrices for a speaker-independent split across the four North S\'{a}mi dialects using meanstd pooling with XLS-R (left) and HuBERT (right) and DNN for classification.}
  \label{fig:conf_matrix}
\end{figure}

\subsection{North S\'{a}mi datasets}

\begin{table}[tbp]
\footnotesize %scriptsize, footnotesize, small, normalsize
\centering
\caption{\textbf{Utterance} counts of the North S\'{a}mi recordings in their respective varieties based on the majority language.}
\vspace{-2mm}
\label{tab:utterances}
\begin{tabular}{ *{8}{c}} %{p{2cm}|p{1.5cm}|p{1.5cm}|p{1.5cm}}
\hline
\multirow{2}{2em}{\textit{Dialect}} & \multicolumn{2}{c}{\textbf{Finnish}} & \multicolumn{2}{c}{\textbf{Norwegian}} & \multicolumn{2}{c}{\textbf{Swedish}} & \multicolumn{1}{c}{\textit{Total}} \\
  & \textit{f} & \textit{m} & \textit{f} & \textit{m} & \textit{f} & \textit{m} & \textit{/dialect}\\
\hline
 \textit{WF} & 21 & 38 & 4564 & 4943 & 0 & 0 & 9566\\
 \textit{EF} & 1861 & 2864 & 835 & 3644 & 0 & 0 & 9204\\
 \textit{SS} & 102 & 212 & 1398 & 2201 & 0 & 0 & 3913\\
 \textit{TS} & 363 & 639 & 624 & 1602 & 63 & 166 & 3457\\
\hline
\textit{Total} & 2347 & 3753 & 7421 & 12390 & 63 & 166 & \textbf{26140}
\end{tabular}
\vspace{-4mm}
\end{table}

%\begin{table}[tbp]
%\footnotesize %scriptsize, footnotesize, small, normalsize
%\centering
%\caption{\textbf{Speaker} counts of the North S\'{a}mi recordings in their respective varieties for each four datasets. Abbreviations for the dialect groups: WF = Western Finnmark, EF = Eastern Finnmark, SS = Sea S\'{a}mi, TS = Torne S\'{a}mi. Abbreviations for the datasets: LIA = Lia S\'{a}pmi, Giellagas = Giellagas Corpus of Spoken Saami Languages, TTS = The closed-source Text-to-speech corpus of North S\'{a}mi, DS = The DigiSami corpus.}
%\vspace{-2mm}
%\label{tab:layerresults}
%\begin{tabular}{ *{10}{c}} %{p{2cm}|p{1.5cm}|p{1.5cm}|p{1.5cm}}
%\hline
%\multirow{2}{2em}{\textit{Dialect}} & \multicolumn{2}{c}{\textbf{LIA}} %& \multicolumn{2}{c}{\textbf{Giellagas}} & \multicolumn{2}{c}{\textbf{TTS}} & \multicolumn{2}{c}{\textbf{DS}} & \multicolumn{1}{c}{\textit{Total}} \\
%  & \textit{f} & \textit{m} & \textit{f} & \textit{m} & \textit{f} & \textit{m} & \textit{f} & \textit{m} & \textit{/dialect}\\
%\hline
% \textit{WF} & 9 & 20 & 1 & 1 & 1 & 1 & 5 & 3 & 41\\
% \textit{EF} & 10 & 28 & 1 & 1 & 0 & 0 & 14 & 9 & 63\\
% \textit{SS} & 7 & 12 & 1 & 3 & 0 & 0 & 0 & 0 & 23\\
% \textit{TS} & 2 & 15 & 2 & 1 & 0 & 0 & 0 & 0 & 20\\
%\hline
%\textit{Total} & 28 & 75 & 5 & 6 & 1 & 1 & 19 & 12 & \textbf{147}\\
%\end{tabular}
%\vspace{+1mm}
%\end{table}
%Looked at the view categories table and removed speakers with less than 4 utterances, got: 147 

Our materials for this paper is combined from \textbf{four} different datasets. 
% TODO: maybe saying something about the indigenous people's right etc?
% The Sámi people have been heavily burdened with modern western research for a long time. So we respect the data usage restrictions and handle it carefully. We encourage using existing datasets instead of recording new ones.
The largest of these is the \textit{LIA S\'{a}pmi North S\'{a}mi corpus} from the National Library of Norway\footnote{\url{https://www.nb.no/sprakbanken/en/resource-catalogue/oai-tekstlab-uio-no-lia-sapmi/}}. This spontaneous speech corpus was recorded and collected during 1960--1990, consisting of all four dialects\footnote{Speaker distribution per dialect: WF--29; EF--38; SS--29; TS--17.} and is openly available online through a user interface: \url{https://tekstlab.uio.no/glossa2/saami}. For a downloadable version, a special agreement is required between the LIA S\'{a}pmi maintainers and the data user(s).

The second dataset was the \textit{The Giellagas Corpus of Spoken Saami Languages}. The North S\'{a}mi part of the corpus consists of spontaneous speech (interviews) from 12 speakers, representing all of the four dialects\footnote{Speakers per dialect: WF--2; EF--2; SS--4; TS--3.}. More information on the corpus is found at: \url{https://www.oulu.fi/en/giellagas-corpus-spoken-saami-languages}.

%and the data itself will be made available through \textit{Kielipankki}, the Language Bank of Finland\footnote{organised by The FIN-CLARIN consortium and The Department of Modern Languages of University of Helsinki. The FIN-CLARIN consortium is a part of the international CLARIN infrastructure and it offers a service for the researchers in Finland for an easy access to all the European CLARIN-compatible language resources.}.

% \url{https://metashare.csc.fi/repository/browse/giellagas-corpus-of-spoken-saami-languages/48f40ef20dd211e5947c005056be118efb5deb1649004ba68dae316a8981b9e3/} % DOESNT WORK PROPERLY IN TEXT

The third and fourth dataset in our materials consist of read speech instead of spontaneous speech. The third dataset was produced as closed source for a North S\'{a}mi Text-to-Speech project with two speakers representing the WF dialect and is not publicly available yet. The fourth dataset was produced by the DigiSami project (\url{https://blogs.helsinki.fi/digisami-project/} and is described in detail in \cite{jokinen2014open,jokinen2014community,jokinen2019researching}. This dataset contains the two Finnmark dialects only (WF -- 8 spkrs; EF -- 23 spkrs). 

%For our experiments, we only used the read speech part of the DigiSami corpus. Information on the availability of the data can be found from these publications as well.

We are unable to release a full public copy of the full combined dataset due to the licence restrictions but have included detailed information to enable comparison to similar datasets and tasks. For some of the described datasets, the data can be requested by contacting the data owners/responsible persons. 

The four datasets were all uniformly pre-processed and split to utterances and labeled according to the dialect groups of the speakers as well as speakers' gender and country of origin (i. e. the majority language presumed to have most influence on their North S\'{a}mi speech). This resulted to altogether \textbf{26140} utterances, the distribution of which is shown in Table~\ref{tab:utterances}. 

% DigiSami ConvSpeech only?: \url{https://etsin.fairdata.fi/dataset/fe5e67ba-fc36-449d-b83e-e1e286cc9dda}; \url{https://blogs.helsinki.fi/digisami-project/publications/}; 

%For all datasets, the dialectal group used here was defined according to the geographical location of the village and the dialectal borders shown in \cite{aikio2022north} and accordingly in Figure~\ref{fig:dialects}. 

%For example in the LIA S\'{a}pmi dataset, no dialectal groups were given in the original metadata but the name of the village was indicated for each recording. Each village was then located by the authors from the dialectal map mentioned above.

Our material consists of 70\% spontaneous speech and 30\% read speech which makes it unbalanced. Also, the majority of the materials were from speakers recorded in Norway (76\% of the combined materials),  23\% in Finland and only 0.8\% in Sweden. According to the utterance distribution counts (Table~\ref{tab:utterances}), 37\% of the materials represented the Western Finnmark dialect (WF), 35\% Eastern Finnmark (EF) -- making them nearly equally represented in the material. For the two other groups, the Sea S\'{a}mi (SS) and the Torne S\'{a}mi (TS), the percentages from the material were only 15\% and 13\%, respectively. In summary, ~70\% of the material represents one of the two Finnmark dialects and 30\% the other two dialect groups, SS and TS. 62\% of the materials were recorded from male speakers and 38\% from female speakers.

Finally, the locations of the recording sites and/or places of origin of the speakers in our corpus are indicated in Figure~\ref{fig:dialects}. Even though the village of An\'{a}r is considered to be outside the traditional S\'{a}mi speaking area\footnote{It has to also noted that in the An\'{a}r area, two other S\'{a}mi languages, Inari S\'{a}mi and Skolt S\'{a}mi are actively spoken.}, there is a lively speaking community there and in this paper, it is presumed that the speakers from An\'{a}r in our material represent the EF dialect group.

%as well as heard from the recordings based on the typical dialectal characteristics that most speakers from An\'{a}r in our material represent the EF dialect group.

\vspace{-2mm}
\subsection{Experimental setup}

To evaluate our data we perform experiments in both speaker-independent (SI) and speaker-dependent (SD) setups. Our aim is to understand whether the model generalizes to unseen speakers and also to examine how much of the performance can be attributed to speakers' idiosyncrasies. For each setup (SI/SD), we further generate 3 data splits and average the results. In particular, for SD case we vary the random seed and perform uniform sampling in the entire dataset with a ratio of 80\% training and 20\% for testing. For SI, in each split we leave 3 speakers from each dialect out (a total of 12 unseen speakers in the test set). In both cases, for validation we split the training set randomly with a 90\%-10\% train-validation ratio.

We run our experiments on all features and splits and report the average accuracies in Table~\ref{tab:classificationresults} -- note that we did not include standard deviations due to space constraints. As some features inherently contain characteristics of the speakers, we speaker normalized the data only for experiments run for eGeMAPS and FBs. For all other features, the data were fed directly to the network following their extraction. In general, our aim is to remove as far as possible as much of the speaker variation in order to better observe the dialectal differences.

For the majority language experiment we run the classifier on all utterances in our data and report the $n$-best reults. In practice, for each $n$, we get the top $n$ results from the classifier (in this case, the $n$ top languages that the classifier finds closest to the given utterance; obtained from the sorted posterior) and report those as a percentage over the actual number of majority language instances for the given category.

%Spont speech: 18289 utt; 70\%
%Read speech: 7851 utt; 30\%

% What about percentages per dataset per dialect? E.g. in the Acapela data, there was only WF and LIA Sápmi had very many.

\begin{table}[tbp]
\small %scriptsize, footnotesize, small, normalsize
\centering
\caption{Unweighted accuracy (\%) across dialects with speaker-dependent (SD) and speaker-independent (SI) splits. Results are averaged across runs. Values in bold are the highest scores per row per SI/SD.}
\vspace{-2mm}
\label{tab:classificationresults}
\begin{tabular}{ *{7}{c}} %{p{2cm}|p{1.5cm}|p{1.5cm}|p{1.5cm}}
\hline
\multirow{2}{3em}{\textit{Feature}} & \multicolumn{2}{c}{\textbf{mean}} & \multicolumn{2}{c}{\textbf{std}} & \multicolumn{2}{c}{\textbf{meanstd}} \\
  & \textit{SD} & \textit{SI} & \textit{SD} & \textit{SI} & \textit{SD} & \textit{SI} \\
\hline
 \textit{eGeMAPS} & 60.1 & \textbf{32.9} & 55.8 & 30.1 & \textbf{63.4} & 32.1\\
 \textit{MFCCs} & 52.9 & 29.7 & \textbf{86.1} & \textbf{38.3} & 83.7 & 37.0\\
 \textit{FBs} & 63.4 & 29.3 & 58.0 & 28.5 & \textbf{67.2} & \textbf{32.7}\\
 \textit{HuBERT} & \textbf{82.5} & 46.2 & 65.1 & 44.5 & 82.2 & \textbf{47.9}\\
 \textit{WavLM} & \textbf{72.7} & 41.0 & 65.9 & 42.6 & 70.6 & \textbf{44.0}\\
 \textit{XLS-R} & \textbf{95.0} & 62.8 & 88.5 & 56.9 & 90.1 & \textbf{62.9}\\
\hline
\end{tabular}
\vspace{-2mm}
\end{table}

\begingroup
\setlength{\tabcolsep}{2pt} % Default value: 6pt
\begin{table}[t]
\small %scriptsize, footnotesize, small, normalsize
\centering
\caption{Majority language influence based on $n$-best language identification results. All reported values are percentages.}
\vspace{-2mm}
\label{tab:majoritynbest}
\begin{tabular}{ *{10}{c}} %{p{2cm}|p{1.5cm}|p{1.5cm}|p{1.5cm}}
\hline
\multirow{2}{4em}{\textit{Majority language}} & \multicolumn{3}{c}{\textbf{1-best}} & \multicolumn{3}{c}{\textbf{2-best}} & \multicolumn{3}{c}{\textbf{5-best}} \\
  & \textit{fi} & \textit{no} & \textit{sv} & \textit{fi} & \textit{no} & \textit{sv} & \textit{fi} & \textit{no} & \textit{sv} \\
\hline
 \textit{Finnish} & \textbf{28.9} & 12.9 & 7.7 & \textbf{39.2} & 30.2 & 14.2 & \textbf{56.3} & 51.5 & 32.3\\
 \textit{Norwegian} & 10.3 & \textbf{20.3} & 13.4 & 22.6 & \textbf{36.8} & 24.4 & 44.5 & \textbf{62.9} & 50.1\\
 \textit{Swedish} & 13.5 & \textbf{17.5} & 17.0 & 32.8 & \textbf{36.7} & 24.5 & \textbf{56.3} & 55.46 & 53.7\\
\hline
\end{tabular}
\vspace{-6mm}
\end{table}
\endgroup

\vspace{-2mm}
\section{Results and Discussion}

An overview of our results can be seen in Tables~\ref{tab:classificationresults} and \ref{tab:majoritynbest}. For dialect classification we obtain the best result for the XLS-R model and meanstd pooling reaching 62.9\% accuracy for SI and 95 \% for SD. For majority language influence we observe a consistent effect of the country where the NS dialects are spoken.

\vspace{-2mm}
\subsection{Classification performance}
We obtain the best classification results for SSL representations (for SI and meanstd, XLS-R 62.9\%, and HuBERT 47.9\%) whereas traditional acoustic features seem to perform systematically with lower accuracy (for SI and meanstd, MFCCs 37\% and FBs 32.7\%) -- note that the random baseline for the data is 25\%. SD splits demonstrate much higher performance than SI splits. This is primarily due to the model learning the idiosyncrasies of the speakers and likely relying its prediction based on speaker rather than the dialect spoken -- due to the uniform sampling, SD splits contain the same speakers in training and testing. Moreover, it is possible that the model can pick fine-grained spectral differences that have to do with the recording conditions. Overall, the SI splits provide a good indication on how separable the dialects are based on the acoustic data. As there were 12 unseen speakers in each SI split, the model has likely learned the segmental and prosodic characteristics of the dialects and is able to make predictions on new speakers. It is interesting to note that mean and meanstd performance is quite close but less so with the std across all features. 
%The study examined the accuracy of the XLS-R method in predicting the dialects of four Sámi dialect groups. 

The results in Figure~\ref{fig:conf_matrix} show that the prediction performance varied between the groups, with WF having the highest at 78.1\%, followed by SS at 71.2\%. TS had the lowest precision at 58.3\%, and EF had 64.2\%. Surprisingly, some samples from TS and EF were predicted as SS, which could be due to the majority language influence (Norwegian) spoken by the speakers in the data, especially since not all of these groups (EF and TS) were geographically close (see Figure~\ref{fig:dialects}). It can also be noted that the prosodic characteristics of the Scandinavian/Germanic languages, Norwegian and Swedish, may contribute to greater differences between the groups spoken in Finland and those spoken in Norway and Sweden.

\vspace{-2mm}
\subsection{Majority language influence}
The state language where the dialects are spoken has an important impact on the NS dialects. For Finnish and Norwegian the LID task identifies the correct language ($1$-best) in 28.9\% and 20.3\% of the cases respectively. For Finnish, if we combine the $1$-best predictions with those of Estonian, two languages that belong to the same language family and are phonetically very close, the LID performance goes up to 40.4\%. If we do the same for $2$-best and $5$-best results we get 51\% and 71.4\%. On the other hand, the results that we get for Swedish are not consistent with the area spoken. The LID task identifies the Swedish NS recordings primarily as Norwegian (17.5\%) and then as Swedish (17\%). Although the difference is small, this might be due to the insufficient data points that we had in our data for Swedish (229 recordings), thus failing to observe more general trends in the data.

\vspace{-2mm}
\section{Conclusions}
In this work we presented an extensive analysis of the classification potential of four NS dialects based on speech. The experiments presented are one of the first evaluating several self-supervised models in DID and showing their performance in the task. We also presented an approach for evaluating the majority language influence on the dialects by running a language identification experiment. We show good prerformance in classifying the dialects using XLS-R and we also provide strong indications of the majority language influence on the dialects. In future work, we will include more NS data in our experiments and fine-tune the XLS-R model with NS data. In addition we plan to use the available textual data in a fusion approach combining embeddings from speech and language models.

\section{Acknowledgements}

\ifinterspeechfinal
   S.K. was supported by the Academy of Finland project no. 340125 “Computational Modeling of Prosody in Speech” and K.H. by The Divvun group at UiT The Arctic University of Norway. The authors wish to acknowledge CSC – IT Center for Science, Finland, for providing the computational resources.
\else
   This will be added later.   
\fi

\pagebreak

\bibliographystyle{IEEEtran}
\bibliography{mybib}

\begin{thebibliography}{10}
\providecommand{\url}[1]{#1}
\csname url@samestyle\endcsname
\providecommand{\newblock}{\relax}
\providecommand{\bibinfo}[2]{#2}
\providecommand{\BIBentrySTDinterwordspacing}{\spaceskip=0pt\relax}
\providecommand{\BIBentryALTinterwordstretchfactor}{4}
\providecommand{\BIBentryALTinterwordspacing}{\spaceskip=\fontdimen2\font plus
\BIBentryALTinterwordstretchfactor\fontdimen3\font minus
  \fontdimen4\font\relax}
\providecommand{\BIBforeignlanguage}[2]{{%
\expandafter\ifx\csname l@#1\endcsname\relax
\typeout{** WARNING: IEEEtran.bst: No hyphenation pattern has been}%
\typeout{** loaded for the language `#1'. Using the pattern for}%
\typeout{** the default language instead.}%
\else
\language=\csname l@#1\endcsname
\fi
#2}}
\providecommand{\BIBdecl}{\relax}
\BIBdecl

\bibitem{chiang2017cross}
C.-Y. Chiang, ``Cross-dialect adaptation framework for constructing prosodic
  models for chinese dialect text-to-speech systems,'' \emph{IEEE/ACM
  Transactions on Audio, Speech, and Language Processing}, vol.~26, no.~1, pp.
  108--121, 2017.

\bibitem{weninger2019deep}
F.~Weninger, Y.~Sun, J.~Park, D.~Willett, and P.~Zhan, ``Deep learning based
  mandarin accent identification for accent robust asr.'' in
  \emph{INTERSPEECH}, 2019, pp. 510--514.

\bibitem{liu2011systematic}
G.~A. Liu and J.~H. Hansen, ``A systematic strategy for robust automatic
  dialect identification,'' in \emph{2011 19th European Signal Processing
  Conference}.\hskip 1em plus 0.5em minus 0.4em\relax IEEE, 2011, pp.
  2138--2141.

\bibitem{jauhiainen2022italian}
T.~Jauhiainen, H.~Jauhiainen, and K.~Lind{\'e}n, ``Italian language and dialect
  identification and regional french variety detection using adaptive naive
  bayes,'' in \emph{Proceedings of the Ninth Workshop on NLP for Similar
  Languages, Varieties and Dialects}.\hskip 1em plus 0.5em minus 0.4em\relax
  The Association for Computational Linguistics, 2022.

\bibitem{kakouros2020dialect}
S.~Kakouros, K.~Hiovain, M.~Vainio, and J.~{\v{S}}imko, ``Dialect
  identification of spoken {North Sámi} language varieties using prosodic
  features,'' \emph{arXiv preprint arXiv:2003.10183}, 2020.

\bibitem{imaizumi2022end}
R.~Imaizumi, R.~Masumura, S.~Shiota, H.~Kiya \emph{et~al.}, ``End-to-end
  japanese multi-dialect speech recognition and dialect identification with
  multi-task learning,'' \emph{APSIPA Transactions on Signal and Information
  Processing}, vol.~11, no.~1, 2022.

\bibitem{zaidan2014arabic}
O.~F. Zaidan and C.~Callison-Burch, ``Arabic dialect identification,''
  \emph{Computational Linguistics}, vol.~40, no.~1, pp. 171--202, 2014.

\bibitem{wang2021end}
D.~Wang, S.~Ye, X.~Hu, S.~Li, and X.~Xu, ``An end-to-end dialect identification
  system with transfer learning from a multilingual automatic speech
  recognition model.'' in \emph{Interspeech}, 2021, pp. 3266--3270.

\bibitem{ali2016automatic}
A.~Ali, N.~Dehak, P.~Cardinal, S.~Khurana, S.~H. Yella, J.~Glass, P.~Bell, and
  S.~Renals, ``Automatic dialect detection in arabic broadcast speech,'' in
  \emph{Interspeech 2016}, 2016, pp. 2934--2938.

\bibitem{Shon2018}
\BIBentryALTinterwordspacing
S.~Shon, A.~Ali, and J.~Glass, ``Convolutional neural network and language
  embeddings for end-to-end dialect recognition,'' in \emph{Proc. Odyssey 2018
  The Speaker and Language Recognition Workshop}, 2018, pp. 98--104. [Online].
  Available: \url{http://dx.doi.org/10.21437/Odyssey.2018-14}
\BIBentrySTDinterwordspacing

\bibitem{simko2018mumbling}
J.~Simko, R.~von Waldenfels, M.~Daniel, N.~Dobrushina, A.~Rabus, A.~S. Suni,
  K.~Hiovain, and M.~T. Vainio, ``Mumbling through a wall: Clustering slavic
  dialects using hierarchical statistical modeling of prosody,'' 2018.

\bibitem{suni2019comparative}
A.~Suni, M.~Wlodarczak, M.~Vainio, and J.~Simko, ``Comparative analysis of
  prosodic characteristics using wavenet embeddings,'' in \emph{20th Annual
  Conference of the International Speech Communication Association (INTERSPEECH
  2019)}.\hskip 1em plus 0.5em minus 0.4em\relax ISCA, 2019.

\bibitem{oord2016wavenet}
A.~v.~d. Oord, S.~Dieleman, H.~Zen, K.~Simonyan, O.~Vinyals, A.~Graves,
  N.~Kalchbrenner, A.~Senior, and K.~Kavukcuoglu, ``Wavenet: A generative model
  for raw audio,'' \emph{arXiv preprint arXiv:1609.03499}, 2016.

\bibitem{gaarding1977scandinavian}
E.~G{\aa}rding, ``The scandinavian word accents,'' \emph{Working papers/Lund
  University, Department of Linguistics and Phonetics}, vol.~8, 1977.

\bibitem{aikio2022north}
A.~Aikio and J.~Ylikoski, ``{North} {Saami},'' in \emph{The Oxford Guide to the
  Uralic Languages}, M.~Bakr{\'o}-Nagy, J.~Laakso, and E.~Skribnik, Eds.\hskip
  1em plus 0.5em minus 0.4em\relax Oxford University Press, 2022, pp. 147--177.

\bibitem{palismaa2001gielas}
M.~Palismaa and I.~M.~G. Eira, ``Gielas gillii, mielas millii 9 --
  {Davvis{\'a}megiela} suopmanat (from language to language, from mind to mind
  9 -- {The} dialects of {North Sami}),'' \emph{Davvi Girji,
  K{\'a}r{\'a}{\v{s}}johka}, 2001.

\bibitem{sammallahti1998saami}
P.~Sammallahti, \emph{The {S}aami languages: an introduction}.\hskip 1em plus
  0.5em minus 0.4em\relax Davvi girji, 1998.

\bibitem{aikio2015variation}
A.~Aikio, L.~Arola, and N.~Kunnas, ``Variation in {North} {Saami},'' in
  \emph{Globalising sociolinguistics: {C}hallenging and expanding theory},
  D.~Smakman and P.~Heinrich, Eds.\hskip 1em plus 0.5em minus 0.4em\relax
  Routledge, 2015, pp. 243--255.

\bibitem{kakouros2018comparison}
S.~Kakouros, O.~R{\"a}s{\"a}nen, and P.~Alku, ``Comparison of spectral tilt
  measures for sentence prominence in speech—effects of dimensionality and
  adverse noise conditions,'' \emph{Speech Communication}, vol. 103, pp.
  11--26, 2018.

\bibitem{kakouros2017evaluation}
------, ``Evaluation of spectral tilt measures for sentence prominence under
  different noise conditions.'' in \emph{Interspeech}, 2017, pp. 3211--3215.

\bibitem{hiovain2018mapping}
K.~Hiovain, A.~S. Suni, J.~\v{S}imko, M.~T. Vainio \emph{et~al.}, ``Mapping
  areal variation and majority language influence in {North S{\'a}mi} using
  hierarchical prosodic analysis,'' in \emph{Proceedings of the 9th
  International Conference on Speech Prosody 2018}.\hskip 1em plus 0.5em minus
  0.4em\relax International Speech Communications Association, 2018.

\bibitem{hiovain2020comparative}
K.~Hiovain, A.~Suni, S.~Kakouros, and J.~{\v{S}}imko, ``Comparative analysis of
  majority language influence on {North} {S{\'a}mi} prosody using
  {WaveNet}-based modeling,'' \emph{Language and Speech}, 2020.

\bibitem{jokinen2016variation}
K.~Jokinen, T.~N. Trong, and V.~Hautam{\"a}ki, ``Variation in spoken {North
  Sami} language.'' in \emph{Interspeech}, 2016, pp. 3299--3303.

\bibitem{eyben2015geneva}
F.~Eyben, K.~R. Scherer, B.~W. Schuller, J.~Sundberg, E.~Andr{\'e}, C.~Busso,
  L.~Y. Devillers, J.~Epps, P.~Laukka, S.~S. Narayanan \emph{et~al.}, ``The
  geneva minimalistic acoustic parameter set (gemaps) for voice research and
  affective computing,'' \emph{IEEE transactions on affective computing},
  vol.~7, no.~2, pp. 190--202, 2015.

\bibitem{povey2011kaldi}
D.~Povey, A.~Ghoshal, G.~Boulianne, L.~Burget, O.~Glembek, N.~Goel,
  M.~Hannemann, P.~Motlicek, Y.~Qian, P.~Schwarz \emph{et~al.}, ``The kaldi
  speech recognition toolkit,'' in \emph{IEEE 2011 workshop on automatic speech
  recognition and understanding}, no. CONF.\hskip 1em plus 0.5em minus
  0.4em\relax IEEE Signal Processing Society, 2011.

\bibitem{hsu2021hubert}
W.-N. Hsu, B.~Bolte, Y.-H.~H. Tsai, K.~Lakhotia, R.~Salakhutdinov, and
  A.~Mohamed, ``Hubert: Self-supervised speech representation learning by
  masked prediction of hidden units,'' \emph{IEEE/ACM Transactions on Audio,
  Speech, and Language Processing}, vol.~29, pp. 3451--3460, 2021.

\bibitem{chen2022wavlm}
S.~Chen, C.~Wang, Z.~Chen, Y.~Wu, S.~Liu, Z.~Chen, J.~Li, N.~Kanda,
  T.~Yoshioka, X.~Xiao \emph{et~al.}, ``Wavlm: Large-scale self-supervised
  pre-training for full stack speech processing,'' \emph{IEEE Journal of
  Selected Topics in Signal Processing}, vol.~16, no.~6, pp. 1505--1518, 2022.

\bibitem{babu2021xls}
A.~Babu, C.~Wang, A.~Tjandra, K.~Lakhotia, Q.~Xu, N.~Goyal, K.~Singh, P.~von
  Platen, Y.~Saraf, J.~Pino \emph{et~al.}, ``Xls-r: Self-supervised
  cross-lingual speech representation learning at scale,'' \emph{arXiv preprint
  arXiv:2111.09296}, 2021.

\bibitem{valk2021voxlingua107}
J.~Valk and T.~Alum{\"a}e, ``Voxlingua107: a dataset for spoken language
  recognition,'' in \emph{2021 IEEE Spoken Language Technology Workshop
  (SLT)}.\hskip 1em plus 0.5em minus 0.4em\relax IEEE, 2021, pp. 652--658.

\bibitem{speechbrain}
M.~Ravanelli, T.~Parcollet, P.~Plantinga, A.~Rouhe, S.~Cornell, L.~Lugosch,
  C.~Subakan, N.~Dawalatabad, A.~Heba, J.~Zhong, J.-C. Chou, S.-L. Yeh, S.-W.
  Fu, C.-F. Liao, E.~Rastorgueva, F.~Grondin, W.~Aris, H.~Na, Y.~Gao, R.~D.
  Mori, and Y.~Bengio, ``{SpeechBrain}: A general-purpose speech toolkit,''
  2021, arXiv:2106.04624.

\bibitem{jokinen2014open}
K.~Jokinen, ``Open-domain interaction and online content in the {Sami}
  language.'' in \emph{LREC}, 2014, pp. 517--522.

\bibitem{jokinen2014community}
K.~Jokinen and G.~Wilcock, ``Community-based resource building and data
  collection,'' in \emph{Spoken language technologies for under-resourced
  languages}, 2014.

\bibitem{jokinen2019researching}
P.~K. Jokinen, ``Researching less-resourced languages: the {Digisami} corpus,''
  in \emph{Proceedings of the Eleventh International Conference on Language
  Resources and Evaluation (LREC-2018)}.\hskip 1em plus 0.5em minus 0.4em\relax
  European Languages Resources Association (ELRA), 2019.

\end{thebibliography}

\end{document}